\documentclass{Interspeech2024}
\usepackage{array}
\usepackage{multirow}
\usepackage{amsmath}
\usepackage{mathrsfs}
\renewcommand{\arraystretch}{1.0}
\newcommand{\modelfont}{\fontfamily{lmtt}\selectfont} %\uppercase

\interspeechcameraready 

\title{Speaker- and Text-Independent Estimation of Articulatory Movements and Phoneme Alignments from Speech}

\name[affiliation={1,2}]{Tobias}{Weise}
\name[affiliation={2}]{Philipp}{Klumpp}
\name[affiliation={1}]{Kubilay}{Can Demir}
\name[affiliation={2,4}]{Paula Andrea}{Pérez-Toro}
\name[affiliation={3}]{Maria}{Schuster}
\name[affiliation={2}]{Elmar}{Noeth}
\name[affiliation={2}]{Bjoern}{Heismann}
\name[affiliation={2}]{Andreas}{Maier}
\name[affiliation={1}]{Seung Hee}{Yang}
\address{
  \small $^1$Speech \& Language Processing Lab. Friedrich-Alexander-Universit\"at Erlangen-N\"urnberg, Germany\\
  \small $^2$Pattern Recognition Lab. Friedrich-Alexander-Universit\"at Erlangen-N\"urnberg, Germany\\
  \small $^3$Department of Otorhinolaryngology, Head and Neck Surgery. Ludwig-Maximilians University, Munich, Germany\\
  \small $^4$GITA Lab, Faculty of Engineering. Universidad de Antioquia, Medell\'in, Colombia}
\email{tobias.weise@fau.de}

\keywords{speech inversion, attention, phoneme alignment, wav2vec 2.0, HPRC, tract variables, multi-task learning}

\begin{document}

\maketitle

 % %%%%%%%%%%%%%%%%%
 % ABSTRACT
\begin{abstract}
This paper introduces a novel combination of two tasks, previously treated separately: acoustic-to-articulatory speech inversion (AAI) and phoneme-to-articulatory (PTA) motion estimation. We refer to this joint task as acoustic phoneme-to-articulatory speech inversion (APTAI) and explore two different approaches, both working speaker- and text-independently during inference.
We use a multi-task learning setup, with the end-to-end goal of taking raw speech as input and estimating the corresponding articulatory movements, phoneme sequence, and phoneme alignment. While both proposed approaches share these same requirements, they differ in their way of achieving phoneme-related predictions: one is based on frame classification, the other on a two-staged training procedure and forced alignment.
We reach competitive performance of $0.73$ mean correlation for the AAI task and achieve up to approximately $\qty{87}{\%}$ frame overlap compared to a state-of-the-art text-dependent phoneme force aligner.
\end{abstract}

% %%%%%%%%%%%%%%%%%
% 1. Introduction
% %%%%%%%%%%%%%%%%%
\section{Introduction}
In phonetics, articulatory configurations are analyzed to understand how different sounds are produced and how they can be classified into phonemes within a particular language's phonological system. Articulators refer to the various parts of the vocal tract and other structures (e.g. tongue, lips, palate) involved in the production of sounds. They are typically measured by placing sensor coils, in a procedure called electromagnetic articulography (EMA), and tracking the position and movement over time during speech. These sensor coordinates are naturally speaker-specific since they depend on the particular vocal tract anatomy of the recorded speaker. Tract Variables (TVs), introduced by Brownman et. al. \cite{browman1990gestural}, on the other hand, combine multiple individual vocal tract articulator movements, that achieve a specific linguistic objective, into defined \textit{gestures} relevant to articulation. Transformations were introduced by Ji \cite{ji2014speaker} to convert EMA sensor coordinates into TVs, which were shown to be less speaker dependent \cite{mcgowan1994recovering} than the original measurements.
\begin{figure}[!htpb]
    \vspace{1.5em}
    \centering
    \includegraphics[scale=0.7]{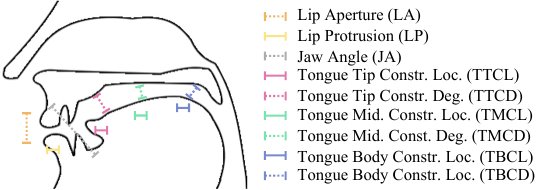}
    \caption{Nine tract variables (TVs), used for speaker-independent articulatory speech inversion. Adapted from \cite{chartier2018encoding,wu2023speakerindependent}.}
    \label{fig:tvs_skull}
\end{figure}

The problem of inverting an original speech signal back to its articulator positions is referred to as acoustic-to-articulatory speech inversion (AAI), which can involve TVs or EMA coordinates as targets. This task has been studied speaker-dependent and speaker-independently in literature: multi-task learning (MTL) \cite{wang2022acoustic, siriwardena2022acoustic}, generative adversarial networks \cite{beguvs2023articulation}, the application to dysarthric speech \cite{maharana2021acoustic}, and speech therapy \cite{benway23c_interspeech, Haldin2021rehabilitation, Triona2020Randomized}, the incorporation of fundamental frequency \cite{siriwardena2023secret}, and others \cite{seneviratne19_interspeech, sivaraman2019unsupervised, shahrebabaki2020sequence} have been explored. A related but less studied problem is taking a sequence of phonemes and mapping it to articulator movements (PTA): gated bidirectional recurrent neural networks \cite{biasutto2018phoneme}, attempts to model the entire vocal tract \cite{ribeiro2022automatic}, comparative studies \cite{abhayjeet2020comparative}, and feed-forward transformers \cite{udupa2021estimating} have been applied, where the latter authors also applied it to AAI in a speaker-dependent setting.
%Model 1 architecture
\begin{figure}
  \centering
  \includegraphics[scale=0.80]{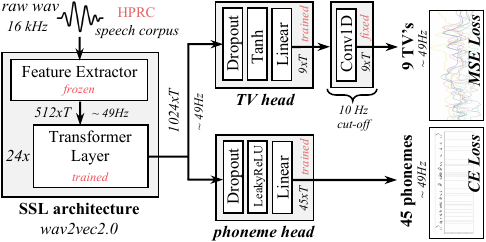}
  \caption{Proposed {\modelfont APTAI} model, based on wav2vec2.0 fine-tuning via frame-classification and TV regression.}
  \label{fig:model1_architecture}
\end{figure}

Phoneme recognition can be described as taking an audio signal as input and producing the corresponding frame-asynchronous phoneme sequence. However, the frame-synchronous relation \cite{li2021combining} is required for the task of phoneme alignment \cite{badlani2022one, zhu2022phone, shih2021rad}, boundary detection, and segmentation \cite{kreuk2020interspeech}. This paper focuses on phoneme recognition and subsequent alignment to the individual frames, which can be beneficial e.g. during speech therapy \cite{li2023improving, jiachen2023unconstrained}. Here, we explore frame-wise classification and forced alignment. Our upper bound is a state-of-the-art (SOTA) text-dependent force aligner. This system relies on both audio and transcriptions as input, which are converted from graphemes to phonemes.% before force aligning them to the input audio sequence. 

This paper introduces APTAI, a novel combination of AAI and PTA in combination with phoneme recognition and alignment. We require that resulting models predict end-to-end (in a therapeutic context) while working speaker- and text-independently during inference. To this end, two different approaches are explored, with Figure \ref{fig:tvs_skull} illustrating the TV regression targets to model articulation.

% %%%%%%%%%%%%%%%%%
% 2. Proposed Approach
% %%%%%%%%%%%%%%%%%
\section{Proposed Approach(es)} \label{sec:approaches}
This paper introduces two approaches, sharing the same requirements outlined in the last paragraph of the introduction. Both make use of MTL optimization, composed of articulator movement regression and phoneme prediction paired with alignment. The main difference is the way they deal with the phoneme-related objective: {\modelfont APTAI} is based on frame classification, whereas {\modelfont f-APTAI} utilizes forced alignment during a two-staged training procedure. Our code is available online\footnote{\url{https://github.com/tobwei/APTAI}}.

Both approaches make use of self-supervised learning (SSL) models but in different setups. Taking ASR as an example, SOTA performance has been achieved using this paradigm, which includes pre-training on large amounts of unlabeled data and fine-tuning on a smaller, labeled dataset relevant to the desired downstream task. We chose {\modelfont wav2vec2} \cite{baevski2020wav2vec}, which optimizes a contrastive loss during pre-training to learn a finite set of speech representations. These can be fine-tuned for a broad set of applications, with ASR as the original intended use case. Thus, such embeddings are expected to capture meaningful features of speech that are relevant for phonemes, which in turn can be identified by specific articulator configurations.

% %%%%%%%%%%%%%%%%%
% Fine-Tune Tabel
\begingroup
\renewcommand{\arraystretch}{1.0}
\begin{table}[h]
\caption{Fine-tuned phoneme recognizer results (PER $[\%]\downarrow$), using CP train/dev splits, for different pre-trained models.}
\vspace{-0.5em}
    \centering
    \scalebox{0.7}{
        \begin{tabular}{cccc}
            \toprule
            {\modelfont wav2vec2-} & CP--test & HPRC--N & HPRC--F\\
            \midrule
            %\hline
            %\multicolumn{1}{l}{mono-lingual, {\modelfont wav2vec2-}} & & & \\
            %\midrule
            %\hline
            {\modelfont base-960h} & $17.77$ & $10.10$ & $19.98$ \\
            {\modelfont large-960h} & $18.71$ & $11.47$ & $24.27$ \\
            {\modelfont large-lv60} & $9.75$ & $4.96$ & $13.76$ \\
            {\modelfont large-960h-lv60} & $9.30$ & $4.55$ & $10.69$ \\
            {\modelfont large-robust} & \textbf{8.83} & \textbf{4.45} & \textbf{10.53} \\
            % \midrule
            %\hline
            %\multicolumn{1}{l}{multi-lingual, {\modelfont wav2vec2-}} & & & \\
            %\midrule
            %\hline
            {\modelfont xls-r-300m} & $11.70$ & $7.77$ & $19.38$ \\
            {\modelfont xls-r-1b} & $18.50$ & $12.69$ & $27.29$ \\
            {\modelfont large-xlsr-53} & $10.17$ & $5.40$ & $14.55$ \\
            % \midrule
            %\hline
            %\multicolumn{1}{l}{mono-lingual, {\modelfont wavlm-}} & & & \\
            %\midrule
            % \hline
            % {\modelfont base} & $16.79$ & $14.46$ & $26.77$\\
            % {\modelfont base-plus} & $11.84$ & $7.64$ & $20.26$\\
            % {\modelfont large} & $16.98$ & $10.83$ & $22.83$\\
            \bottomrule
            %\hline
        \end{tabular}
    }
    \label{tab:phoneme_recognizer_fine_tuning_results}
\end{table}
\endgroup

% %%%%%%%%%%%%%%%%%
% Frame-Classification: APTAI
\subsection{Frame Classification: {\modelfont APTAI}} \label{sec:aptai}
Of the two proposed approaches, {\modelfont APTAI} follows a more classical setup, refer to Figure \ref{fig:model1_architecture} for an overview. The general idea is to fine-tune {\modelfont wav2vec2} to make use of its pre-trained speech representations, which is the reason why we keep the feature extractor frozen (pre-trained weights), and only train the transformer layers (pre-trained initialization) in addition to two added heads (randomly initialized). Furthermore, we add a convolutional layer (fixed parameters), which behaves like a low-pass (sinc) filter, adapted from \cite{parrot2019independent}. This enforces the smoothness of the predicted TV trajectories, which is required since frame-based signal regression typically suffers from high-frequency noise between the individual frame predictions.

An $\qty{16}{kHz}$ input speech signal $x(t)$ is divided into $T$ frames $\bm{x}_t \in \mathbb{R}^{512}$ at $\qty{49}{Hz}$ by the feature encoder. After passing the transformer layers, producing $\bm{h}_t \in \mathbb{R}^{1024}$, the TV head takes this output and ultimately predicts $\bm{\hat{y}}^{tv}_{t} \in \mathbb{R}^{TV}$ smoothed $TV = 9$ values for each frame $t$. As part of the MTL goal, this head optimizes the reconstruction mean square error (MSE) loss between the predicted $\bm{\hat{y}}^{tv}_{t}$ and ground truth $\bm{y}^{tv}_{t}$ TV values, which is expressed in the second term of Equation \ref{eq:L_FC}. The phoneme head also takes $\bm{h}_t$ as input and predicts a probability distribution $\hat{p}_{t,c}$ over $\mathcal{C} = 45$ phoneme labels per frame $t$, with $c \in \mathcal{C}$. This frame-wise classification is optimized via cross-entropy (CE) loss between the predicted $\hat{p}_{t,c}$ and ground truth ${p}_{t,c}$ probability distribution (see first term in Equation \ref{eq:L_FC}). Applying $softmax$ to the resulting logits and choosing the phoneme label $c$ that yields the maximum probability per frame $t$ will result in an alignment, whilst a phoneme sequence can be obtained by grouping over the individual frame predictions. Finally, Equation \ref{eq:L_FC} shows the MTL loss $\mathcal{L_{\text{FC}}}$ for the {\modelfont APTAI} approach, with $\lambda$ as weighting factor.
\vspace{-0.5em}
\begin{equation} \label{eq:L_FC}
\mathcal{L_{\text{FC}}} = - \frac{1}{T} \sum_{t=1}^{T} \sum_{c=1}^{\mathcal{C}} p_{t,c} \log(\hat{p}_{t,c}) + \lambda \frac{1}{T} \sum_{t=1}^{T} (\bm{y}^{tv}_{t} - \bm{\hat{y}}^{tv}_{t})^2
\end{equation}

% %%%%%%%%%%%%%%%%%
% forced-alignment: f-APTAI
\subsection{Forced Alignment: {\modelfont f-APTAI}}
\begin{figure}
  \centering
  \includegraphics[scale=0.7]{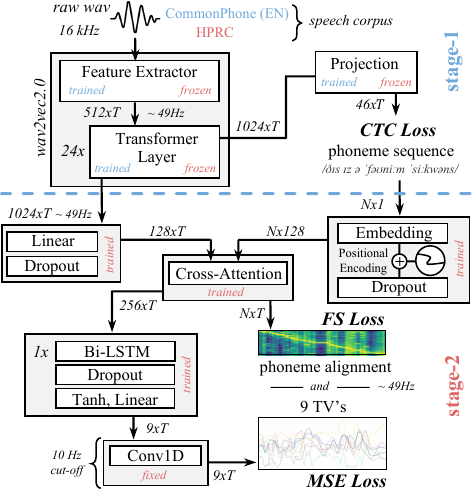}
  \vspace{-0.5em}
  \caption{Proposed {\modelfont f-APTAI} model, based on TV regression and a two-staged forced alignment via cross-attention.}
  \label{fig:model2_architecture}
\end{figure}

The idea behind the second approach {\modelfont f-APTAI} is to make use of hidden representations from a fine-tuned phoneme recognizer in combination with a forced alignment of the predicted output phoneme sequence. To this end, we use a two-staged approach during training, depicted in Figure \ref{fig:model2_architecture}. We make use of different datasets for the two stages, more details in section \ref{sec:datasets}.

For the first stage, we fine-tune the same SSL architecture ({\modelfont wav2vec2}) used in {\modelfont APTAI}, by adding a linear layer producing $\bm{l}_{t} \in \mathbb{R}^{\mathcal{C}_{\emptyset}}$ representing the same $\mathcal{C} = 45$ phoneme labels with the addition of a blank token $\emptyset$, per frame $t \in T$ (see section \ref{sec:aptai}). Similar to the ASR application, we optimize this model using the connectionist temporal classification (CTC) loss. This optimization behaves like a state machine, similar to hidden markov models (HMM), and only requires a phoneme sequence as additional input during training. However, CTC does not produce an alignment but rather outputs a frame-asynchronous (in our case) phoneme label sequence through a frame-synchronous decoding procedure (beam search), utilizing the blank token and multiple possible alignment paths. Given a true phoneme label sequence $\mathcal{W}$, then $\mathcal{S}$ represents all possible paths that map from $\mathcal{W}$ to $T$ by removing repeated labels and blanks. Then, $P(s_t \mid \bm{l}_t)$ represents the output of the model at $t$ by applying $softmax$ to $\bm{l}_t$, with $[s_{1:T}] \in \mathcal{S}$. Adapted from \cite{li2021combining}, the CTC loss can be defined as:
\vspace{-0.5em}
\begin{equation} \label{eq:ctc}
\mathcal{L_{\text{CTC}}} = - \log \sum_{\mathcal{S}} \prod_{t=1}^{T} P(s_t \mid \bm{l}_t)
\end{equation}
\vspace{-0.5em}

\begin{figure}[!htpb]
  \centering
  \includegraphics[width=\linewidth]{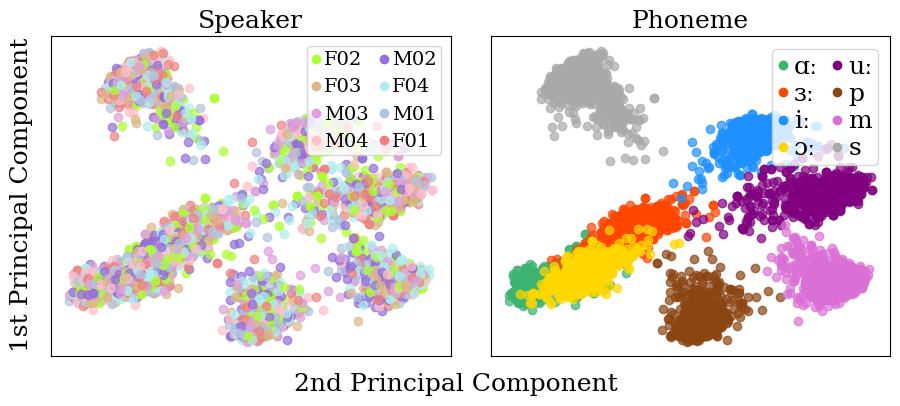}
  \vspace{-2em}
  \caption{PCA of the embeddings from the best-performing fine-tuned phoneme recognition model from Table \ref{tab:phoneme_recognizer_fine_tuning_results}.}
  \label{fig:embeddings} 
\end{figure}

The second stage of {\modelfont f-APTAI} incorporates the frozen model trained during stage-1. Specifically, two parts are extracted and used during training of stage-2: the predicted CTC-based phoneme sequence (upper bound for stage-2) and the output of the last transformer layer. Here, let the former be $[p_{1:N}] \in \mathcal{P}^{N}$, where $p_n \in \mathcal{C}$, and $N$ the maximum sequence length. The last transformer layer output can be expressed as matrix $\bm{H}$, consisting of $\bm{h}_t \in \mathbb{R}^{1024}$ column vectors, with $t \in T$. This can be understood as acoustic phoneme embeddings since the stage-1 objective (see Equation \ref{eq:ctc}) led to accordingly optimized weights. A principal component analysis (PCA) of these embeddings (extracted from the HPRC--N dataset, see section \ref{sec:datasets}) can be seen in Figure \ref{fig:embeddings}. The setup is similar to \cite{tom2022wav2vec} and shows good speaker independence with phoneme clustering of exemplary chosen elongated vowels, a fricative, nasal, and plosive. The performed neural forced alignment is inspired by \cite{zhu2022phone} and has the goal of producing a monotonic alignment, such that it aligns each phoneme label $p_n$ to a subset of consecutive hidden frame representations $\bm{h}_t$. Therefore, one of the MTL optimization goals of {\modelfont f-APTAI} is to learn a matrix $\bm{A} \in \mathbb{R}^{NxT}$ that aligns $\mathcal{P}^N$ to $\bm{H}$. This objective is centered around a cross-attention computation between a learned linear projection of $\bm{h}_t$ to $\bm{h}^{p}_t \in \mathbb{R}^{128}$ resulting in $\bm{H}_{p} \in \mathbb{R}^{Tx128}$, and a learned embedding of $\mathcal{P}^{N}$. This embedding is created via projection of each $p_n \in \mathcal{P}^{N}$ to $\mathbb{R}^{128}$ and the addition of a sinusoidal positional encoding \cite{vaswani2017attention}, ultimately resulting in matrix $\bm{P} \in \mathbb{R}^{128xN}$. Finally, the cross-attention layer computes the alignment matrix $\bm{A} = softmax(\bm{H}_p \cdot \bm{P})$. We constrain $\bm{A}$ to be monotonic and diagonal, which is inspired by the forward-sum (FS) loss used in HMM systems, and adapted from \cite{badlani2022one, shih2021rad}. See the first term in Equation \ref{ep:fAPTAI}, where $\mathcal{O}$ is the optimal alignment.
\vspace{-0.75em}
\begin{equation} \label{ep:fAPTAI}
\mathcal{L_{\text{FA}}} = - \sum_{\bm{H}_p, \bm{P} \in \mathcal{O}} \log P(\bm{P} \mid \bm{H}_p) + \lambda \frac{1}{T} \sum_{t=1}^{T} (\bm{y}^{tv}_{t} - \bm{\hat{y}}^{tv}_{t})^2
\end{equation}
Additionally, the cross-attention layer produces a hidden representation matrix $\in \mathbb{R}^{256 \times T}$. This sequence of column vectors over $T$ frames serves as input for the TV regression part of the {\modelfont f-APTAI} model. Initially, it is passed through a single bi-directional long short-term memory (LSTM) layer, the output of which is ultimately projected to $\mathbb{R}^{TV}$. Moreover, the same fixed-parameter convolutional low-pass (sinc) filter as in {\modelfont APTAI} is used to ensure the prediction of smooth TV trajectories $\bm{\hat{y}}^{tv}_{t}$. Consequently, the same MSE loss is also optimized, see the second term in Equation \ref{ep:fAPTAI}.
% %%%%%%%%%%%%%%%%%
% Model 2 architecture

% %%%%%%%%%%%%%%%%%
% 3. Experimental Setup
% %%%%%%%%%%%%%%%%%
\section{Experimental Setup} \label{sec:experiments}
It should be noted that our upper bound for both approaches, in terms of phoneme recognition and alignment, is a SOTA \cite{zhu2022phone} text-dependent force aligner from WebMAUS \cite{kisler2017multilingual}. The reason for this is that we produce our ground truth phoneme labels and time steps via this web API. We make use of CommonPhone (see section \ref{sec:datasets}) for its robustness and this dataset utilized the same process, so we apply the same to HPRC, the second dataset that we use to guarantee compatibility. 
% %%%%%%%%%%%%%%%%%
% Embedding Figure

\subsection{Datasets} \label{sec:datasets}
One of the two datasets that we use during experiments is Common Phone (CP) \cite{klumpp2022common}, which is based on the crowd-sourced Common Voice \cite{ardila2019common}. Here, we utilize the English subset (45 phoneme labels). The main motivation behind using CP is that we want to build a robust system. When comparing CP to e.g. TIMIT \cite{garofolo1993timit}, this robustness becomes evident: one is recorded in the same acoustically controlled environment with professional equipment, and the other is based on recordings from people's smartphones in many different uncontrolled environments. 

\begingroup
\renewcommand{\arraystretch}{1}
\begin{table}[]
\caption{Leave-one-speaker-out results (mean and deviation across eight test speakers) for the two proposed approaches.}
\vspace{-0.5em}
    \centering
    \scalebox{0.69}{
        \begin{tabular}{l cc cc}
            \toprule
            %& \multicolumn{2}{c}{\textit{TV metrics}} & \multicolumn{2}{c}{\textit{PHN metrics}} \\
            Model, Test Data&PCC$\uparrow$ & RSME$[mm]\downarrow$ & PER$[\%]\downarrow$ & Overlap$[\%]\uparrow$ \\
            \midrule
            %\multicolumn{5}{c}{\textbf{\modelfont APTAI}} \\
            %\midrule
            {\modelfont APTAI}, HPRC--N&\textbf{0.73} $\pm$ 0.03&\textbf{0.67} $\pm$ 0.03&6.25 $\pm$ 1.30&\textbf{87.38} $\pm$ 1.16\\
            {\modelfont APTAI}, HPRC--F&0.69 $\pm$ 0.03&0.72 $\pm$ 0.03&6.41 $\pm$ 1.76&84.91 $\pm$ 1.93\\

            \midrule
            %\multicolumn{5}{c}{\textbf{\modelfont f-APTAI}} \\
            %\midrule
            {\modelfont f-APTAI}, HPRC--N&0.71 $\pm$ 0.03&0.68 $\pm$ 0.03&\textbf{4.36} $\pm$ 0.07&76.18 $\pm$ 1.59\\
            {\modelfont f-APTAI}, HPRC--F&0.65 $\pm$ 0.03&0.74 $\pm$ 0.03&10.29 $\pm$ 3.62&72.93 $\pm$ 2.92\\
            \bottomrule
        \end{tabular}
    }
    \label{tab:results_HPRC}
\end{table}
\endgroup

The second dataset we use contains articulator-related information in the form of EMA sensor data. This dataset is the Haskins Production Rate Comparison (HPRC) \cite{tiede2017quantifying}, which contains recordings from four female and four male subjects reciting 720 phonetically balanced IEEE sentences at "normal" (HPRC--N) and "fast" (HPRC--F) speaking rates. The speakers in this dataset repeat utterances, however, we randomly select only one repetition per utterance and speaker. Furthermore, we used the MAUS aligner to create our ground truth phoneme labels and time steps. This dataset comes with labels from another aligner, but we wanted to make it compatible with the CP dataset. Next, we performed pre-processing on the EMA data: some of the coordinates contained \textit{NaN} values, where we applied linear interpolation to remedy this problem before low-pass (Butterworth) filtering the sensor data with $\qty{20}{Hz}$ to eliminate recording related noise. After this, the EMA coordinates were transformed into nine TVs (see Figure \ref{fig:tvs_skull}) and some final processing was applied to them. The original EMA data was sampled at $\qty{100}{Hz}$, resulting in TVs at the same rate. We resampled them to $\qty{49}{Hz}$ to synchronize them with the output frame rate of {\modelfont wav2vec2}. Finally, we applied utterance-wise z-score normalization based on the individual TVs. 

%%%%%%%%%%%%%%%%
% Alignment Figure
\begin{figure}[t]
  \centering
  \includegraphics[width=\linewidth]{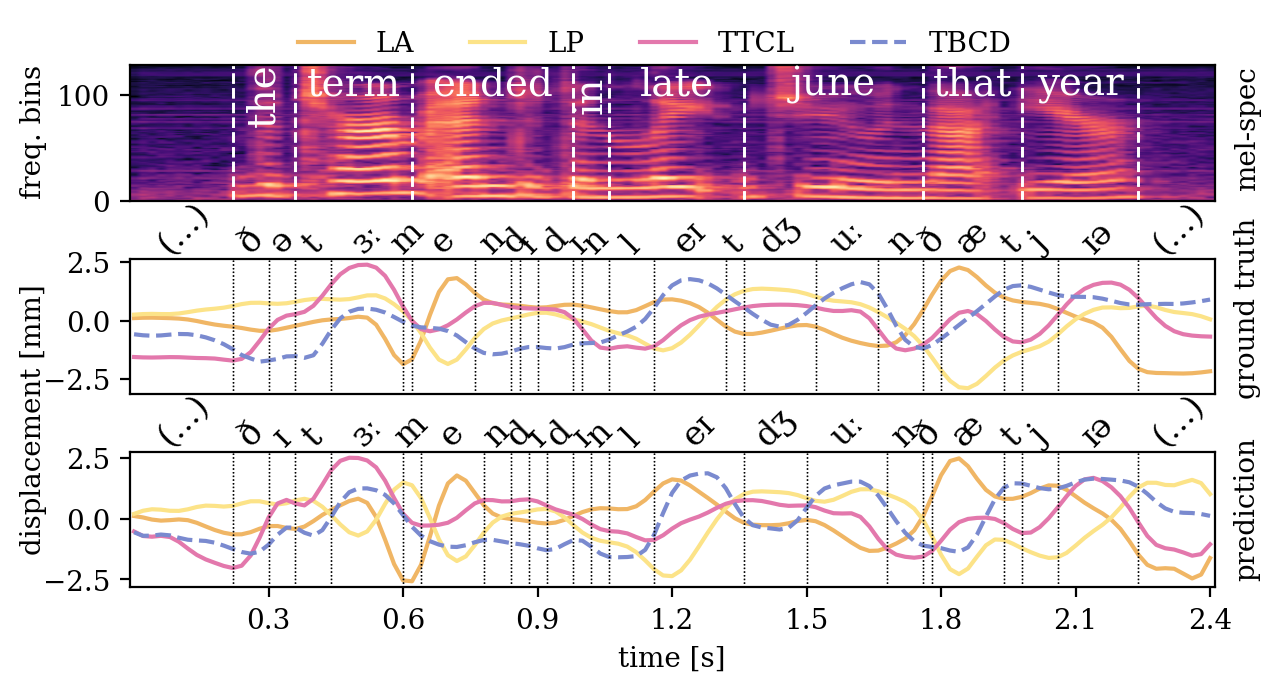}
  \vspace{-2em}
  \caption{Example model prediction ({\modelfont APTAI}) and ground truth for an unseen speaker (only a selection of all TVs is shown).}
  \label{fig:alignment}
\end{figure} 

%%%%%%%%%%%%%%%%
% Model evaluation
\subsection{Model Evaluation} \label{sec:evaluation}
We evaluate the APTAI task in terms of the two MTL sub-objectives. The articulation regression performance is evaluated using two well-known metrics: the root mean square error (RMSE) based on the normalized values and the Pearson correlation coefficient (PCC). To evaluate the phoneme recognition and alignment performance, we use the phoneme error rate (PER), where the ground truth is based on the webMAUS grapheme-to-phoneme conversion. Phoneme alignment is also evaluated regarding this text-dependent upper bound, using the frame-wise overlap (percentage of correctly predicted frames). 

%%%%%%%%%%%%%%%%
% Model training
\subsection{Model Training} \label{sec:training}
The following setup was used to train/validate our two proposed approaches, using the PyTorch framework. For CP, we used the official train/dev/test splits. To test the performance of our models, we used HPRC. Here, we applied leave-one-speaker-out testing, i.e., data from seven speakers was used for training/validation (90\%/10\%), and the data of the remaining speaker was used to test (separated by speaking rates). Additionally, we performed the training split in such a way that only unseen utterances were used for validation. The same optimizer (Adam), learning rate ($1\mathrm{e}{-5}$), learning-rate scheduler (warm-up, static, and decaying epochs), batch size of 5, and model selection metric (TV RMSE) were used for both proposed approaches. We experimented with MTL strategies (e.g. alternating epochs) but with no improvement in performance. 

{\modelfont APTAI}, utilizing {\modelfont wav2vec2-large-robust} (see Table \ref{tab:phoneme_recognizer_fine_tuning_results}), was trained for 20 epochs, with 20\% dropout, and combined HPRC--N and --F for training/validation. In terms of the MTL loss optimization, we set $\lambda = 1$ thus weighting both tasks equally, which resulted in the best performance.

Fine-tuning of the phoneme recognizer for stage-1 of {\modelfont f-APTAI} was based on {\modelfont wav2vec2-large-robust} (best performance, see Table \ref{tab:phoneme_recognizer_fine_tuning_results}) with a batch size of 2, 160 epochs, learning rate of $5\mathrm{e}{-6}$, a final dropout of 10\%, and model selection based on validation PER. For stage-2, we trained for 60 epochs, used only HPRC--N (since including F would negatively impact the PER of stage-1), set $\lambda = 0.4$, and $N = 60$, with shorter phoneme sequences being padded. Finally, the implementation of the FS loss was taken from \cite{shih2021rad}.

 % %%%%%%%%%%%%%%%%%
 % 4. Results and Discussion
 % %%%%%%%%%%%%%%%%%
\section{Results and Discussion}
Table \ref{tab:phoneme_recognizer_fine_tuning_results} reveals that CP is a noisy dataset, while HPRC is not. This results in better PER for "normal" speaking rates, while "fast" are more challenging (also for human listeners), with {\modelfont wav2vec2-large-robust} performing best.

\begin{figure}[t]
    \centering
    \includegraphics[width=\linewidth]{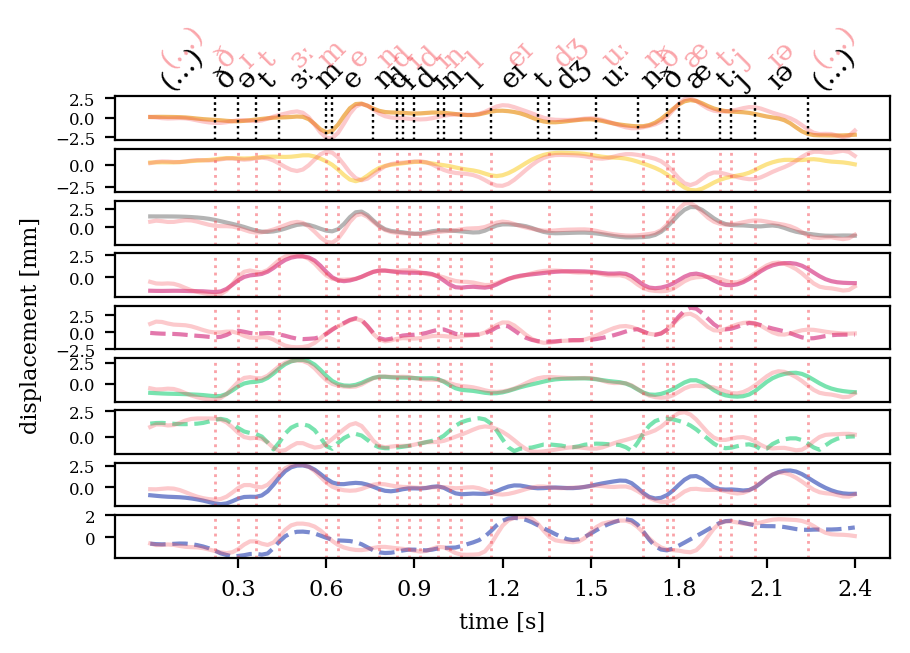}
    \vspace{-2.2em}
    \caption{Ground truth and model prediction ({\modelfont APTAI}) in red color, of an unseen speaker (refer to Figure \ref{fig:tvs_skull} as legend).}
    \label{fig:9tvs_pred}
\end{figure}

Table \ref{tab:results_HPRC} shows the main evaluation test results of the introduced APTAI task, conducted in a speaker-independent (LOSO) setting. Figure \ref{fig:alignment} illustrates prediction performance, showing a selection of TVs for improved readability, whilst Figure \ref{fig:9tvs_pred} shows all TVs individually. In terms of TV metrics, both models perform similarly, with {\modelfont APTAI} achieving the best mean PCC of $0.73$. Comparing this result to other works is difficult since setups are not uniform (e.g. trimming of silence), and reproduced results do not match originally reported ones \cite{wang2022acoustic, shahrebabaki2020sequence}. However, reported speaker-independent PCC results on HPRC roughly range from $35\%$ to $76\%$, so we achieve competitive performance. In terms of phoneme recognition and alignment, frame classification outperforms the forced alignment approach by $11.20\%$, achieving a frame overlap of $87.38\%$. Shih et. al. \cite{shih2021rad} reported that in their experiments, a wider receptive field lead to alignment instability. The fact that we use hidden transformer representations, capturing weighted global sequence dependencies, might explain the reduced alignment performance, which requires future research.
Overall, the work of Siriwardena et. al. \cite{siriwardena2022acoustic} is similar, however, they report a PER of approx. 27\% (and no alignment metric) since they see the phoneme-related objective as an auxiliary task to improve TV-related performance, while we see both tasks as equally important.

When looking at Table \ref{tab:mean_TV_metrics} and Figure \ref{fig:9tvs_pred}, it is noticeable that especially the regression of TMCD and TBCD perform significantly worse when compared to the other TVs, hampering the overall mean PCC. This needs further investigation since other papers do not seem to suffer from this problem.

%%%%%%%%%%%%%%%%%%%
% 9 TVs Figure

\begingroup
\renewcommand{\arraystretch}{1}
\begin{table}[]
\caption{Individual TV metrics, in terms of mean and deviation across the leave-one-speaker-out experiments ({\modelfont APTAI} model).}
\vspace{-0.5em}
    \centering
    \scalebox{0.7}{
        \begin{tabular}{c cc cc}
            \toprule

            & \multicolumn{2}{c}{\textbf{HPRC--N}} & \multicolumn{2}{c}{\textbf{HPRC--F}} \\
            
            \multirow{2}{*}[1.25em]{\textbf{TV's}} & PCC$\uparrow$ & RSME$[mm]\downarrow$ & PCC$\uparrow$ & RSME$[mm]\downarrow$ \\
            
            \midrule
            {LA} &0.87$\pm$0.03&0.49$\pm$0.06 & 81.76$\pm$4.89 & 0.57$\pm$0.07 \\
            %\midrule
            {LP} &0.75$\pm$0.08&0.66$\pm$0.10 & 66.93$\pm$8.57 & 0.75$\pm$0.10 \\
            %\midrule
            {JA} &0.82$\pm$0.04&0.57$\pm$0.06 & 73.97$\pm$4.19 & 0.67$\pm$0.06 \\
            %\midrule
            {TTCL} &0.84$\pm$0.04&0.54$\pm$0.06 & 81.85$\pm$3.25 & 0.56$\pm$0.05 \\
            %\midrule
            {TTCD} &0.79$\pm$0.04&0.61$\pm$0.06 & 74.14$\pm$5.48 & 0.67$\pm$0.06 \\
            %\midrule
            {TMCL} &0.82$\pm$0.03&0.57$\pm$0.04 & 79.38$\pm$2.47 & 0.60$\pm$0.04 \\
            %\midrule
            {TMCD} &0.37$\pm$0.11&1.07$\pm$0.09 & 27.94$\pm$11.34 & 1.13$\pm$0.09 \\
            %\midrule
            {TBCL} &0.77$\pm$0.04&0.64$\pm$0.05 & 74.36$\pm$4.36 & 0.67$\pm$0.06 \\
            %\midrule
            {TBCD} &0.54$\pm$0.15&0.88$\pm$0.14 & 56.57$\pm$14.53 & 0.85$\pm$0.14 \\
            \bottomrule
        \end{tabular}
    }
    \label{tab:mean_TV_metrics}
\end{table}
\endgroup

% %%%%%%%%%%%%%%%%%
% 5. CONCLUSION
% %%%%%%%%%%%%%%%%%
\section{Conclusion}
This paper introduced APTAI, a novel combination of two tasks previously viewed separately. We investigated two different approaches, sharing the same robust requirements but differing mainly in their method of phoneme prediction and alignment. Here, the frame classification based {\modelfont APTAI} model performed better, especially in terms of phoneme-related metrics. However, {\modelfont f-APTAI}, based on forced alignment, has potentially more room for improvement in future work. An example of this, applicable to both models and requiring new pre-training, is changing the output frame rate of {\modelfont wav2vec2} to $\qty{10}{ms}$ instead of $\qty{20}{ms}$ by changing the stride of the feature extractor, to improve alignment performance \cite{zhu2022phone} and enable $\qty{100}{Hz}$ TV regression.

\newpage % page 5 can be ACKNOWLEDGEMENTS and REFERENCES only!
% %%%%%%%%%%%%%%%%%
% 6. ACKNOWLEDGMENTS
% %%%%%%%%%%%%%%%%%
\section{Acknowledgements}
\ifinterspeechfinal
    We gratefully acknowledge funding for this study by Friedrich-Alexander-University Erlangen-Nuermberg, Medical Valley e.V. and Siemens Healthineers AG within the framework of d.hip campus.
\else
    Suppressed due to anonymous submission to INTERSPEECH 2024.
\fi

% %%%%%%%%%%%%%%%%%
% 6. REFERENCES
% %%%%%%%%%%%%%%%%%
\bibliographystyle{IEEEtran}
\bibliography{mybib}

\end{document}